\documentclass[final,5p,twocolumn,authoryear]{elsarticle}

\usepackage{amssymb}
\usepackage{amsmath}
\usepackage{algpseudocode}
\usepackage{algorithm}
\usepackage{url}

\journal{Astronomy $\&$ Computing}

\begin{document}

\begin{frontmatter}



\title{CLiMB: A Domain-Informed Novelty Detection Clustering Framework for Galactic Archaeology and Scientific Discovery} 


\author[inst1]{Lorenzo Monti\corref{cor1}}
\ead{lorenzo.monti@inaf.it}

\author[inst1]{Tatiana Muraveva}

\author[inst2]{Brian Sheridan}

\author[inst1]{Davide Massari}

\author[inst1]{Alessia Garofalo}

\author[inst1]{Gisella Clementini}

\author[inst2,inst3]{Umberto Michelucci}

\cortext[cor1]{Corresponding author}

\affiliation[inst1]{organization={INAF, Osservatorio di Astrofisica e Scienza dello Spazio},
            addressline={via Piero Gobetti 93/3}, 
            city={Bologna},
            postcode={40129}, 
            country={Italy}}

\affiliation[inst2]{organization={TOELT LLC, Machine Learning Research and Development Department},
            city={Winterthur},
            postcode={8406}, 
            state={Zurich},
            country={Switzerland}}

\affiliation[inst3]{organization={Computer Science Department, Lucerne University of Applied Sciences and Arts},
            city={Luzern},
            postcode={6002}, 
            country={Switzerland}}

\begin{abstract}
In data-driven scientific discovery, a challenge lies in classifying well-characterized phenomena while identifying novel anomalies. Current semi-supervised clustering algorithms do not always fully address this duality, often assuming that supervisory signals are globally representative. Consequently, methods often enforce rigid constraints that suppress unanticipated patterns or require a pre-specified number of clusters, rendering them ineffective for genuine novelty detection. To bridge this gap, we introduce CLiMB (CLustering in Multiphase Boundaries), a domain-informed framework decoupling the exploitation of prior knowledge from the exploration of unknown structures. Using a sequential two-phase approach, CLiMB first anchors known clusters using metric-adaptive constrained partitioning, and subsequently applies density-based clustering to residual data to reveal arbitrary topologies. We demonstrate this framework on RR Lyrae stars data from the {\it Gaia} Data Release 3. CLiMB attains an Adjusted Rand Index of 0.829 with 90\% seed coverage in recovering known Milky Way substructures, outperforming heuristic and constraint-based baselines, which stagnate below 0.20. Furthermore, sensitivity analysis confirms CLiMB’s superior data efficiency, showing monotonic improvement as knowledge increases. Finally, the framework successfully isolates three distinct dynamical features (Shiva, Shakti, and the Galactic Disk) in the unlabelled field, validating its potential for scientific discovery.
\end{abstract}




\end{frontmatter}


\section{Introduction}
\label{sec:intro}
The hierarchical mass assembly of the Milky Way (MW) is a fundamental prediction of the $\Lambda$ Cold Dark Matter ($\Lambda$CDM) model, suggesting that the Galactic halo is a complex repository of debris from past mergers with dwarf galaxies and globular clusters \citep{White1991, Johnston1996}. Identifying these ancient structures is essential for reconstructing the evolutionary history of our Galaxy. While stars from a single progenitor tend to maintain coherence in the space of integrals of motion (IoM), their detection is often hindered by significant overlaps in phase space, high background noise from \textit{in-situ} populations, and the intrinsic sparsity of key stellar tracers \citep{Helmi2000, Dodd2023}.


RR Lyrae stars (RRLs) are pulsating variable stars, serving as ideal tracers of old ($\ge 10$ Gyr), metal-poor stellar populations prevalent in the MW halo. RRLs are excellent standard candles for measuring distances due to their well-defined luminosity-metallicity relations in the visual bands and period-luminosity-metallicity relations in the near- and mid-infrared bands (e.g., \citealt{Longmore1986}; \citealt{Clementini2003}; \citealt{Bono2003}; \citealt{Sollima2008}; \citealt{Muraveva2018}). Additionally, RRLs are effective metallicity tracers, as their metallicity can be estimated from photometric parameters, such as the pulsation period and Fourier decomposition parameters of their light curves, without requiring spectroscopic data (e.g., \citealt{Jurcsik1996}; \citealt{Morgan2007}; \citealt{Muraveva2025}).

As remnants tracers of ancient stellar systems, RRLs are particularly valuable for studying the dynamical history of the Galaxy, including past accretion events that have shaped its halo \citep{massari2019origin, helmi2020streams}. These events, involving the merger of dwarf galaxies or globular clusters, leave kinematic and spatial signatures in the form of stellar streams and substructures, which can be traced using IoM, such as the total energy ($E$), the angular momentum along the $Z$-axis ($L_z$), and the perpendicular component of angular momentum ($L_\perp$) \citep{helmi2000mapping, ceccarelli2024walk, Dodd2023}. The ESA {\it Gaia} mission \citep{Prusti2016} has revolutionized this field, providing high-precision astrometric and spectroscopic data. In this study, we leverage a sample of 4,933 RRLs from {\it Gaia} Data Release 3 (DR3; \citealt{Vallenari2023}), for which full 6D phase-space information (coordinates, proper motions, distances, and radial velocities) is available, to compute IoM and evaluate our clustering framework.


Traditional unsupervised clustering approaches treat all data points as equally unknown, ignoring the vast amount of domain knowledge already accumulated about the Galactic halo. Conversely, supervised classification methods are limited to assigning stars to known groups, inherently failing to discover new accretion events. This tension underscores the importance of developing a framework in modern Galactic archaeology that can precisely recover well-characterized substructures while remaining sensitive to novel, unmapped features.

Alternative approaches, such as metric learning, attempt to reshape the feature space to satisfy constraints \citep{baghshah2010kernel, salunke2012constrained}. While powerful, these methods often struggle with local validity; a distance metric learned from a specific subset of known classes may not be applicable to the geometry of unknown regions. More recently, multi-stage frameworks have been proposed to address these rigidities. For instance, \citet{smieja2020classification} decomposite clustering into binary classification and supervised grouping ($S^3C^2$), while \citet{ienco2018semi} utilise multiresolution autoencoders to capture different granularities. Yet, even within these hybrid frameworks, the explicit decoupling of \textit{known structure recovery} from \textit{novelty discovery} remains under-explored.

The application domain further highlights this gap. While bioinformatics has seen extensive semi-supervised clustering (SSC) application—particularly in gene expression and document clustering \citep{gu2013efficient, jiang2013extracting}, astronomical applications have received minimal explicit attention in the semi-supervised literature \citep{baumann2022algorithm}. This is a significant oversight, as current galactic surveys (e.g., \textit{Gaia}, \citealt{Prusti2016}) produce high-dimensional kinematic data in which known stellar streams coexist with undiscovered accretion remnants.

To address these limitations, we introduce CLiMB (CLustering in Multiphase Boundaries), a domain-informed novelty detection framework. CLiMB is motivated by the observation that separating constraint satisfaction from density-based exploration can resolve the tension between prior knowledge and novel discovery \citep{ienco2022deep}. Unlike traditional methods that enforce global constraints, CLiMB operates via a sequential, two-phase architecture. The first phase, \textit{K-Bound}, utilises customizable distance metrics (such as Euclidean, Mahalanobis, and custom metrics) and different boundaries to anchor clusters strictly to user-provided seed points. The second phase applies density-based clustering to the unassigned residual data, enabling the detection of arbitrary, non-convex topologies without forcing them into the mold of the priors.

We applied the CLiMB framework to the Gaia DR3 catalog of RRLs to dissect the dynamical architecture of the Milky Way (MW) halo. The framework’s two-phase architecture proved robust against the dataset's sparsity, significantly outperforming heuristic baselines like C-DBSCAN \citep{ruiz2007c} and SS-DBSCAN \citep{abdulhameed2024ss}. In the initial constrained phase, CLiMB successfully recovered eight known substructures based on IoM similarity, including the Gaia Sausage/Enceladus (GSE) and Sequoia. Crucially, the exploratory second phase unveiled three previously unassigned dynamical groups in the residual data: the Galactic Disk and the distinct accreted substructures Shiva and Shakti \citep{Malhan2024}. By isolating these novel features from the unlabelled field, features that remain elusive for traditional clustering methods, CLiMB characterized a total of 11 substructures, offering new insights into the MW's assembly history.



The paper is organised as follows: Section~\ref{sec:related_works} reviews the relevant literature in semi-supervised clustering and its applications in Galactic archaeology. Section~\ref{sec:methods} details the mathematical architecture of the CLiMB framework and its two-phase clustering approach. Section~\ref{sec:experiments} describes the experimental setup, including the \textit{Gaia} RRL dataset preparation, the calculation of IoM, and the benchmarking protocol against baseline algorithms. Section~\ref{sec:results} presents the comparative analysis of the results, the sensitivity study on prior knowledge, and the scientific validation of the discovered substructures. Finally, Section~\ref{sec:conclusions} provides a summary of our findings and discusses future directions for novelty detection in large-scale astronomical surveys.

\section{Related Works}
\label{sec:related_works}

Semi-supervised clustering (SSC) constitutes a critical subfield of machine learning that addresses the inherent ill-posedness of unsupervised clustering by incorporating limited supervisory information into the partitioning process. This supervision, typically manifested as class labels for a data subset or as pairwise constraints (i.e., must-link and cannot-link pairs), serves to regularize the problem, guiding the algorithm toward a semantically meaningful solution that aligns with available domain knowledge as described in \citep{basu2005semi, qin2019research}. 

\subsection{Paradigms in Semi-Supervised Clustering}
The primary paradigms for integrating supervision in clustering can be broadly categorized into two major approaches: the direct modification of the clustering algorithm and the transformation of the underlying data representation.
Early work predominantly focused on algorithm-level integration through informed initialization or \textbf{seeding}. In this paradigm, labeled data points are used to define the initial cluster centroids. By commencing the iterative optimization process from a more favorable position in the solution space, such methods mitigate the risk of convergence to suboptimal local minima, enhancing both the accuracy and stability of the final partition \citep{basu2002semi}. This principle is a direct precursor to the initial phase of our proposed framework, although traditional seeding lacks the spatial boundaries necessary for novelty detection.

A more flexible form of algorithmic-level integration is realized through \textbf{constraint-based clustering}. Foundational algorithms like COP-Kmeans \citep{wagstaff2001constrained} and PCKmeans \citep{basu2004active} modify the cluster assignment step to enforce \textit{must-link} and \textit{cannot-link} constraints. This paradigm was later adapted for density-based algorithms, which are inherently better suited for discovering non-globular clusters. For instance, C-DBSCAN \citep{ruiz2007c} first partitions the data into numerous local micro-clusters while strictly respecting \textit{cannot-link} constraints, and only subsequently merges them based on \textit{must-link} constraints and proximity. While robust in enforcing local labels, C-DBSCAN is prone to "chain-reaction" merging in high-density regions, where distinct stellar streams might be incorrectly connected by bridges of background noise.

A distinct approach involves \textbf{heuristic-guided clustering}, which modifies algorithmic behavior not with specific pairwise constraints, but with domain-guided rules. Instead of providing examples, this paradigm injects general rules based on domain knowledge to guide the clustering process. The SS-DBSCAN method, as described in \citet{abdulhameed2024ss}, exemplifies this by altering the core point selection criteria of DBSCAN. In this model, a point can only initiate a new cluster if it is both in a dense region and satisfies an external, user-defined "importance" function. While this allows for the integration of high-level domain knowledge, the effectiveness of such methods is highly dependent on the quality and generalizability of the handcrafted heuristic rule.

Moreover, \textbf{metric learning} aims to learn a distance function—often a Mahalanobis metric—under which the data's geometric structure conforms to the supervisory information \citep{xing2002distance, ye2007adaptive}. While powerful in reshaping the feature space, metric learning often struggles with "local validity"; a distance metric optimized for a specific subset of known classes may be ill-suited for the geometry of unknown regions.

More recently, the field has moved toward \textbf{hybrid multi-stage frameworks} and deep clustering \citep{ren2019semi}. These advanced methods often attempt to decouple the clustering process into distinct stages of supervision-guided partitioning and unconstrained exploration. Yet, as noted by \citet{ienco2022deep}, the tension between exploitation (adhering to prior knowledge) and exploration (discovering intrinsic structure) remains a persistent challenge. CLiMB is designed to resolve this tension by formally decoupling these two objectives into a sequential, multiphase architecture.

\subsection{Clustering in Galactic Archaeology}
In the context of Galactic archaeology, the identification of stellar streams has traditionally relied on density-based unsupervised algorithms, most notably DBSCAN \citep{ester1996DBSCAN} and HDBSCAN \citep{campello2013dHDBSCAN}, due to their ability to detect non-convex shapes and handle background noise. These methods have been instrumental in mapping the MW's accretion history \citep{koppelman2019a, Cabrera2024}. However, purely unsupervised methods are highly sensitive to hyperparameter tuning and often struggle with the "varying density" problem inherent in Galactic surveys, where dense disk populations coexist with sparse halo streams.

To incorporate physical priors, model-dependent algorithms like \textit{STREAMFINDER} \citep{malhan2018streamfinder} utilize stellar evolutionary models and orbital integration to find coherent structures. While physically rigorous, these methods are computationally intensive and potentially biased toward structures that fit the assumed potential. Conversely, purely data-driven approaches like \textit{StarGO} \citep{yuan2018stargo} utilize Self-Organizing Maps (SOMs) to find over-densities in high-dimensional space but lack a formal mechanism to integrate existing catalogs of known structures. Furthermore, while these methods focus on mapping the global manifold, they often lack the fine-grained control needed to isolate specific overlapping substructures during the learning process.

\subsection{Bridging the Gap: Decoupling and Novelty Detection}
The challenge of performing \textit{novelty detection} within a semi-supervised framework remains an open research area. Most density-based SSC variants, such as SS-DBSCAN \citep{lelis2009semi} or the heuristic-guided SS-DBSCAN \citep{abdulhameed2024ss}, attempt to guide cluster formation through density-connected components or domain rules. However, in high-density regions where different dynamical entities overlap, label propagation can lead to "chaining effects," where known labels bleed into unrelated structures.

Recent hybrid frameworks have proposed multi-stage architectures to address these rigidities. For instance, the $S^3C^2$ algorithm decomposes clustering into binary classification and supervised grouping \citep{smieja2020classification}, while other methods utilize multiresolution autoencoders to capture different structural granularities \citep{ienco2018semi}. CLiMB builds upon the observation that separating constraint satisfaction (exploitation) from density-based exploration is crucial for scientific discovery \citep{ienco2022deep}. By utilizing a two-phase architecture, CLiMB avoids the pitfalls of global constraint propagation, ensuring that prior knowledge stabilizes known structures without constraining the algorithm's ability to discover novel, arbitrary topologies in the residual data.

\section{Methods}
\label{sec:methods}

\subsection{CLiMB: A Domain-Informed Novelty Detection Clustering Algorithm}
In this work, we propose CLiMB (CLustering in Multiphase Boundaries), a novel clustering algorithm, called Domain-Informed Novelty Detection Clustering, designed to analyse datasets containing both known, well-characterized components and novelty regions where the discovery of new structures is crucial. Unlike traditional clustering methods that operate exclusively under unsupervised assumptions, our approach integrates prior domain knowledge (via "seed points") with the capability to identify and group unexpected and innovative patterns.

The algorithm aims to overcome an inherent limitation in existing clustering approaches: the inability to effectively leverage partial domain knowledge while maintaining robustness in discovering previously unknown structures. Domain-Informed Novelty Detection Clustering is based on the premise that many real-world datasets contain both well-understood components, which can be guided by constraints, and unexplored regions that require adaptive discovery mechanisms. This is achieved through a sequential, two-phase architecture.

\subsection{Phase 1: K-Bound,  Constrained Clustering}
The first phase, named \textit{K-Bound}, implements a revised version of the K-means \citep{mcqueen1967KMEANS} algorithm enhanced with multiple constraint mechanisms. Given a dataset $\mathbf{X} = \{\mathbf{x}_1, \mathbf{x}_2, \ldots, \mathbf{x}_n\} \subset \mathbb{R}^d$ and a predefined number of constrained clusters $k$, \textit{K-Bound} iteratively refines cluster centroids while enforcing three types of constraints.

\begin{itemize}
    \item \textbf{Density Constraint:} To focus on well-defined regions, a point $\mathbf{x}_i$ is only considered for assignment if its local density $\rho(\mathbf{x}_i)$ is above a specified threshold $\tau_{\rho}$. The local density is calculated using a Gaussian kernel density estimate:
    \[
    \rho(\mathbf{x}_i) = \frac{1}{n} \sum_{j=1}^{n} \exp\left(-\frac{d(\mathbf{x}_i, \mathbf{x}_j)^2}{2\sigma^2}\right)
    \]
    
    where the density of a point $\mathbf{x}_i$ is the sum of contributions from all other points in the dataset, and the contribution of each point $\mathbf{x}_j$ decays smoothly in a Gaussian manner as its distance $d(\mathbf{x}_i, \mathbf{x}_j)$ increases. Kernel bandwidth, or $\sigma$, defines the scale of the neighborhood; a smaller $\sigma$ considers only very close neighbors, while a larger $\sigma$ results in a smoother density profile. A point $\mathbf{x}_i$ is admitted for the constrained clustering phase only if its local density, $\rho(\mathbf{x}_i)$, satisfies the condition $\rho(\mathbf{x}_i) \ge \tau_{\rho}$. Points failing this condition are deferred to the exploratory phase.


    \item \textbf{Distance Constraint:} Following the density filter, a candidate point $\mathbf{x}_i$ is evaluated for cluster assignment based on a strict proximity criterion. The assignment is contingent upon both identifying the nearest centroid and satisfying a maximum distance threshold, $\tau_d$. We first determine the index of the nearest centroid, $j^*$, as:
        \[
        j^* = \operatorname*{arg\,min}_{l \in \{1,\ldots,k\}} d(\mathbf{x}_i, \mathbf{c}_l)
        \]
        The formal assignment rule for the point's label, $L(\mathbf{x}_i)$, is then defined as:
        \[
        L(\mathbf{x}_i) = 
        \begin{cases} 
          j^* & \text{if } d(\mathbf{x}_i, \mathbf{c}_{j^*}) \leq \tau_d \\
          \text{unassigned} & \text{otherwise}
        \end{cases}
        \]
    This constraint effectively imposes a boundary of constant metric radius $\tau_d$ around each centroid, defining the maximum spatial extent of a cluster. Its primary function is to enforce cluster compactness and prevent the inclusion of peripheral points that, while closer to one centroid than any other, may not be representative of the cluster's core structure. This ensures a high degree of intra-cluster affinity. Points that fail to satisfy this criterion, despite potentially residing in dense regions, are considered unassigned and are consequently passed to the exploratory phase for analysis.
    
    \item \textbf{Radial Constraint:} To prevent centroids from drifting excessively from their knowledge-guided initial positions, we enforce a radial constraint with threshold $\tau_r$. The position of each centroid $\mathbf{c}_j$ at iteration $t+1$ is constrained relative to its initial position $\mathbf{c}_j^{(0)}$:
    \[
    \|\mathbf{c}_j^{(t+1)} - \mathbf{c}_j^{(0)}\|_2 \leq \tau_r
    \]

    This condition defines a permissible region of movement for each centroid. If an update calculation yields a position outside this region, the centroid is repositioned to the nearest point on the boundary, effectively capping its displacement. This radial constraint ensures robust convergence and guaranties that the final cluster definitions do not diverge substantially from the prior knowledge supplied to the algorithm.
\end{itemize}

\subsubsection{Advanced Initialization and Metric Flexibility}
\textit{K-Bound}'s effectiveness is enhanced by its sophisticated initialization and distance metric options.

\paragraph{Initialization Strategies:} CLiMB supports multiple seed point mechanisms. Initialization can be random, or it can be guided by a list of initial centroid coordinates. For more granular control, the algorithm accepts a dictionary structure $\mathcal{S} = \{(\mathbf{c}_i, \mathbf{S}_i)\}_{i=1}^k$, where $\mathbf{c}_i$ is the initial centroid coordinate and $\mathbf{S}_i = \{\mathbf{s}_{i,1}, \ldots, \mathbf{s}_{i,m_i}\}$ is a set of associated seed points. The dictionary keys are used as the initial centroid positions, and after the algorithm converges, the points in $\mathbf{S}_i$ are guaranteed to be assigned to cluster $i$, ensuring that prior knowledge is respected.

\paragraph{Metric Flexibility:} CLiMB supports multiple distance metrics for the function $d(\cdot, \cdot)$:
\begin{itemize}
    \item \textbf{Euclidean Distance}: $d(\mathbf{x}_i, \mathbf{x}_j) = \|\mathbf{x}_i - \mathbf{x}_j\|_2$.
    \item \textbf{Mahalanobis Distance}: $d(\mathbf{x}_i, \mathbf{x}_j) = \sqrt{(\mathbf{x}_i - \mathbf{x}_j)^T \mathbf{V}^{-1} (\mathbf{x}_i - \mathbf{x}_j)}$, where $\mathbf{V}$ is the covariance matrix of the data. The use of the Mahalanobis metric is particularly critical when dealing with stellar substructures in the space of integrals of motion. Unlike the Euclidean metric, which assumes isotropic (spherical) cluster shapes, the Mahalanobis distance accounts for the correlations and different scales between features such as energy ($E$) and angular momentum ($L_z$). In Galactic archaeology, accreted structures often appear as elongated or tilted "shards" rather than globular distributions due to the physical constraints of tidal stripping and orbital evolution. By incorporating the covariance matrix, CLiMB can define "ellipsoidal" boundaries that better conform to the natural dynamical manifold of stellar streams, preventing the misclassification of field stars that might be geometrically close in Euclidean space but are dynamically distinct.
    \item \textbf{Custom Metrics}: Users can define application-specific distance functions via a flexible interface.
\end{itemize}

\subsection{Phase 2: Exploratory Novelty Detection}
Points not assigned during the \textit{K-Bound} phase form the set $\mathbf{X}_{\text{unassigned}}$. These points undergo exploratory clustering using density-based methods to discover novel, potentially non-convex patterns. CLiMB implements a strategy pattern, allowing for seamless integration of multiple algorithms, including DBSCAN \citep{ester1996DBSCAN}, HDBSCAN \citep{campello2013dHDBSCAN, campello2015hierarchical}, and OPTICS \citep{ankerst1999optics}.

For the default DBSCAN-based exploration, two parameters are defined: a neighborhood radius $\epsilon$ and a minimum number of points $\text{MinPts}$. A point $\mathbf{x}_i \in \mathbf{X}_{\text{unassigned}}$ is identified as a core point if its $\epsilon$-neighborhood contains at least $\text{MinPts}$ points. Clusters are then formed by connecting core points and their neighbors. Labels assigned in this phase are indexed to be distinct from the constrained cluster labels.

\subsection{Convergence, Complexity, and Implementation}

\subsubsection{Convergence Criteria}
The iterative nature of the K-Bound phase requires a robust set of criteria to determine when a stable and meaningful cluster configuration has been achieved. The process is designed to terminate when the cluster assignments and centroid positions no longer exhibit significant changes. This is governed by two distinct conditions.

The first condition for termination is based on the principle of centroid stability. The iterative refinement process is considered to have converged when the displacement of all centroids between two consecutive iterations becomes negligible. We quantify this by monitoring the maximum displacement observed across all centroids. Specifically, the algorithm halts when the largest distance between a centroid's position at iteration $t$, $\mathbf{c}_j^{(t)}$, and its updated position at iteration $t+1$, $\mathbf{c}_j^{(t+1)}$, falls below a user-defined tolerance threshold, $\tau_c$. This condition is expressed formally as:

\[
\max_{j \in \{1,\ldots,k\}} \|\mathbf{c}_j^{(t+1)} - \mathbf{c}_j^{(t)}\|_2 < \tau_c
\]

This use of the maximum displacement provides a conservative criterion. It ensures that the entire system has stabilized, as even the most mobile centroid has settled. The tolerance parameter $\tau_c$ is a hyperparameter that balances the trade-off between the precision of the final centroid placement and the total computational cost; a smaller $\tau_c$ leads to a more refined solution at the expense of a potentially higher number of iterations.

To guarantee termination under all circumstances, a secondary criterion is implemented. The algorithm will halt if it reaches a predefined maximum number of iterations, \textit{max\_iter}. This condition is crucial to prevent cases of non-convergence, such as oscillations where centroids shift between a finite set of positions in successive iterations without ever meeting the stability criterion. This ensures that the algorithm always terminates in a finite amount of time, providing a deterministic upper bound on its runtime.

\subsubsection{Computational Complexity}
\label{subsec:complexity}
The computational complexity of the CLiMB algorithm is analyzed by considering its two sequential phases. The runtime of the \textit{K-Bound Phase} is characterized as $O(t \cdot n \cdot k \cdot d)$, where $t$ is the number of iterations, $n$ is the number of data points, $k$ is the number of clusters, and $d$ is the feature space dimensionality. This complexity arises from the iterative core of the algorithm. Within each iteration, the primary computational complexity is the \textit{Label Assignment} phase. This step requires computing the distance between each of the $n$ points and all $k$ centroids, resulting in an $n \times k$ distance matrix. This operation has a cost of $O(n \cdot k \cdot d)$. The subsequent \textit{Centroid Update} phase, where new cluster means are calculated, is less costly, with a complexity of $O(n \cdot d)$. As the assignment phase is dominant, it dictates the complexity of a single iteration. Additionally, it is important to note a one-time pre-computation cost of $O(n^2 \cdot d)$ for the initial density estimation, which can be a significant factor for datasets with a large number of samples. Comparing the two phases within an iteration, the Label Assignment ($O(n \cdot k \cdot d)$) is computationally more expensive than the Centroid Update ($O(n \cdot d)$). Therefore, the complexity of a single iteration is governed by the Label Assignment phase.

The complexity of the subsequent \textit{Exploratory Phase} is contingent upon the chosen algorithm and the number of unassigned points, $n' \le n$. The default algorithm, DBSCAN, typically exhibits an average-case complexity of $O(n' \log n')$ when optimized with spatial indexing structures. For the alternative supported algorithms, such as OPTICS and HDBSCAN, the worst-case computational complexity is $O((n')^2)$. While computationally more intensive, these methods provide more sophisticated models for density-based discovery, which can be advantageous for complex data structures.

In most practical applications, the runtime of the K-Bound phase constitutes the dominant portion of the total computational cost, particularly when the number of constrained clusters $k$ is non-trivial and the number of unassigned points $n'$ is significantly smaller than the total sample size $n$.

\subsubsection{Implementation Details}
CLiMB is implemented in Python and is compatible with the \textit{scikit-learn}\footnote{https://scikit-learn.org} ecosystem. The modular architecture allows for easy extension. For flexible parameter configuration, the algorithm employs a builder design pattern. This design facilitates systematic parameter tuning and enhances experimental reproducibility. The package also includes comprehensive 2D and 3D visualization tools. The source code is publicly available under the MIT license and is hosted on GitHub\footnote{https://github.com/LorenzoMonti/CLiMB}, where documentation, tests and examples are also provided. The algorithm's implementation prioritizes computational efficiency by exploiting optimized, vectorized operations provided by the NumPy\footnote{https://numpy.org/} and SciPy\footnote{https://scipy.org/} libraries for all distance calculations and array manipulations. Memory usage is managed by recalculating intermediate data structures, such as the distance matrix, within each iteration, thus avoiding the storage of a complete computational history. The pseudo-code is presented in Algorithm \ref{alg:climb}.

\begin{algorithm}[ht]
\caption{The CLiMB Algorithm}
\label{alg:climb}
\begin{algorithmic}[1]
\State \textbf{Input:} Dataset $\mathbf{X}$, number of clusters $k$, thresholds $\tau_{\rho}, \tau_{d}, \tau_{r}, \tau_{c}$
\Statex \hspace{1.1em} Seed points $\mathcal{S}$ (optional), Exploratory algorithm $A_{expl}$
\vspace{0.5em}
\Statex \textbf{Phase 1: Constrained Clustering (K-Bound)}
\State Initialize centroids $\mathbf{C}^{(0)} = \{\mathbf{c}_1^{(0)}, \ldots, \mathbf{c}_k^{(0)}\}$ using $\mathcal{S}$ or randomly.
\State Compute local density $\rho(\mathbf{x}_i)$ for all $\mathbf{x}_i \in \mathbf{X}$.
\For{$t = 0, \ldots, \text{max\_iter}$}
    \State Assign labels: $L_i \leftarrow \arg\min_j d(\mathbf{x}_i, \mathbf{c}_j^{(t)})$ if $\rho(\mathbf{x}_i) \geq \tau_{\rho}$ and $d(\mathbf{x}_i, \mathbf{c}_j^{(t)}) \leq \tau_{d}$.
    \State Update centroids $\mathbf{C}^{(t+1)}$ based on the mean of assigned points.
    \State Enforce radial constraint: $\|\mathbf{c}_j^{(t+1)} - \mathbf{c}_j^{(0)}\|_2 \leq \tau_r$.
    \If{$\max_j \|\mathbf{c}_j^{(t+1)} - \mathbf{c}_j^{(t)}\|_2 < \tau_c$} \textbf{break} \EndIf
\EndFor
\State Let $\mathbf{X}_{\text{unassigned}}$ be the set of unassigned points.
\vspace{0.5em}
\Statex \textbf{Phase 2: Exploratory Clustering}
\If{$\mathbf{X}_{\text{unassigned}}$ is not empty}
    \State Run $A_{expl}$ on $\mathbf{X}_{\text{unassigned}}$ to get exploratory labels $L_{expl}$.
    \State Combine constrained and exploratory labels.
\EndIf
\vspace{0.5em}
\State \textbf{Output:} Final cluster labels for all points in $\mathbf{X}$.
\end{algorithmic}
\end{algorithm}

\section{Experimental Evaluation}
\label{sec:experiments}
To validate the performance and robustness of our proposed framework, we designed a comprehensive experimental evaluation. Our primary goal is to demonstrate that CLiMB's two-phase, adaptive-metric architecture offers a superior solution for \textit{domain-informed novelty detection clustering} compared to existing semi-supervised methods. The experiments are structured to assess not only the overall clustering quality but also to independently evaluate the distinct contributions of CLiMB's constrained knowledge-recovery phase and its unsupervised exploratory discovery phase

The primary case study is drawn from the domain of Galactic archaeology, a field where identifying sparsely distributed stellar structures within massive, noisy datasets presents a significant challenge perfectly aligned with the problem CLiMB is designed to solve. We benchmark CLiMB against alternative density-based semi-supervised algorithms, evaluating their performance on a real-world dataset of RRLs from the {\it Gaia} DR3. The following sections detail the dataset, the comparative methodology, and the evaluation results.

\subsection{Data Selection: RRLs from Gaia DR3}

\begin{figure*}[t!]
\centering
\includegraphics[width=16cm]{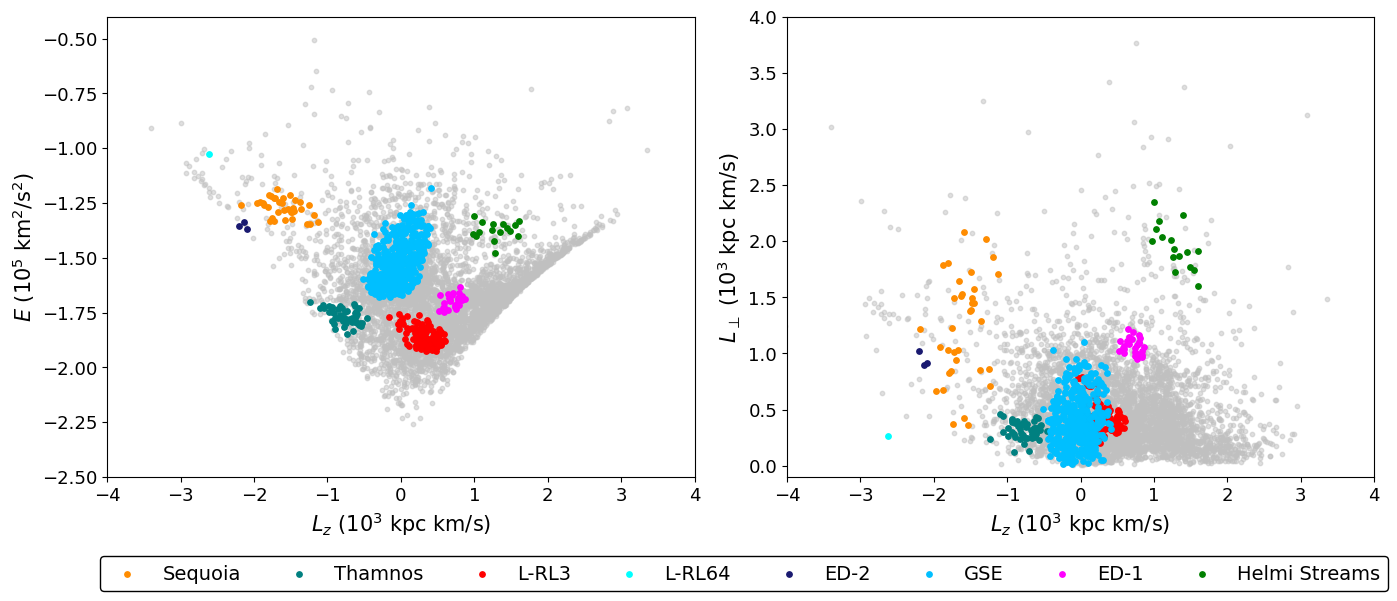}
\includegraphics[width=16cm]{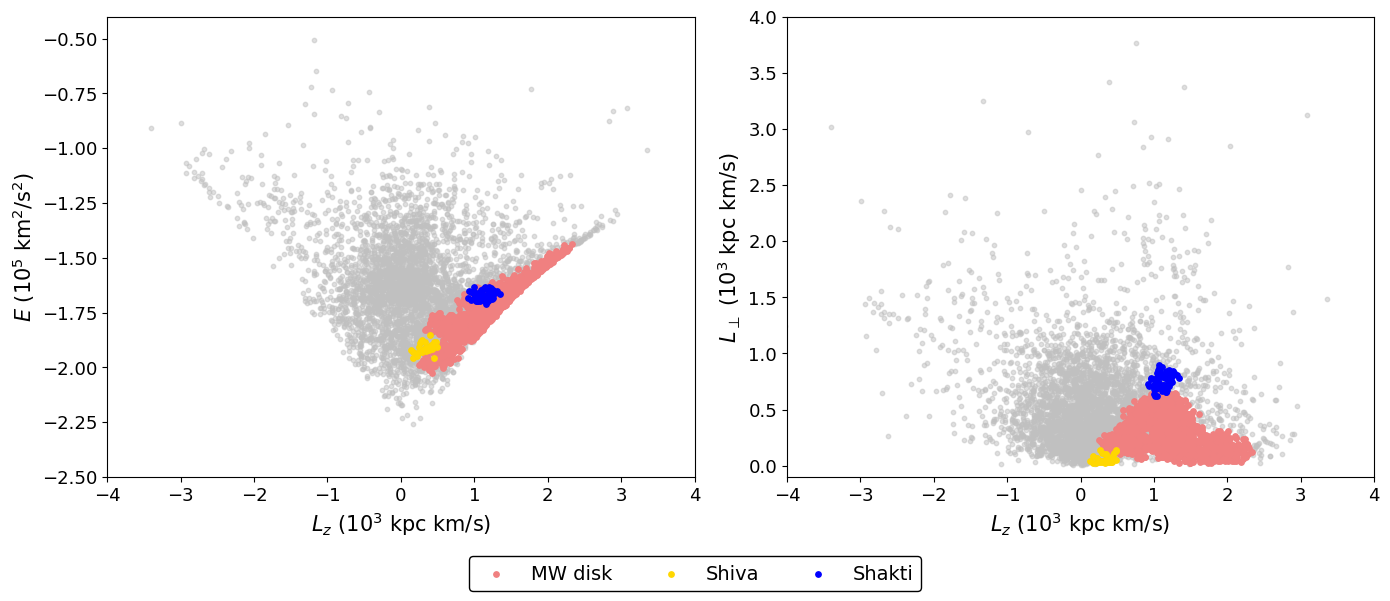}
\caption{Distribution of 4,933 RRLs from {\it Gaia} DR3 in the $E$-$L_z$ and $L_z$-$L_\perp$ planes, colour-coded by the substructure to which they belong. Grey dots indicate RRLs not assigned to any substructure. The top panels show RRLs identified by the CLiMB algorithm in known MW substructures from \citet{Dodd2023}, while the bottom panels show substructures not reported by \citet{Dodd2023}, discovered during the exploratory phase of the CLiMB algorithm.}\label{fig:clusters_all}
\end{figure*}




{\it Gaia} DR3 provides a clean catalogue of 270,891 RRLs, processed via the Specific Object Study pipeline (SOS Cep\&RRL; \citealt{Clementini2023Gaia}). In \citet{Muraveva2025}, we refined this selection to a sample of 258,696 stars, which we consider bona fide RRLs. From this census, we selected stars with available coordinates, radial velocities, and proper motions in the {\it Gaia} Early Data Release 3 (EDR3, \citealt{Brown2021}) and DR3 \citep{Vallenari2023} catalogues. Geometric distances were derived from {\it Gaia} EDR3 parallaxes using a Bayesian approach \citep{BJ2021}. This selection resulted in a final sample of 4,933 RRLs (see Figure~\ref{fig:clusters_all}). For these stars, we calculated the IoM ($E, L_z, L_\perp$) using the AGAMA software package \citep{AGAMA}, adopting the MW potential from \citet{McMillan2017}.

From a machine learning perspective, we project this data into a compact 3-dimensional feature space $\mathbf{X} \in \mathbb{R}^{n \times 3}$:
\begin{enumerate}
    \item \textbf{Angular Momentum ($L_z$):} The component of angular momentum along the Galactic $Z$-axis.
    \item \textbf{Total Energy ($E$):} The orbital energy of the star.
    \item \textbf{Perpendicular Angular Momentum ($L_\perp$):} The component of angular momentum perpendicular to the rotation axis ($L_\perp = \sqrt{L_x^2 + L_y^2}$), which effectively traces the inclination of the orbit.
\end{enumerate}

This feature space is notoriously difficult for clustering due to the varying densities of the streams and the non-Gaussian nature of the background noise.

\subsection{Preprocessing and Feature Engineering}
\label{sec:preprocessing}

To prepare the raw astrophysical data for the semi-supervised pipeline, we implemented a specific preprocessing routine to mitigate biases and ensure isotropic metric properties.

\paragraph{Energy Centering:} To mitigate biases related to the absolute scale of total energy and to facilitate the computation of the Mahalanobis distance, we centered the energy distribution. Specifically, we subtracted the mean background energy level from the original values:

\[
\mathbf{e}' = \mathbf{e} - \mu_e
\]

This procedure ensures that the feature space is centered around the origin regarding the energy component.

\paragraph{Scaling:} Given the heterogeneous ranges of the features ($L_z$ and $L_\perp$ in kpc km/s, $E$ in km$^2$/s$^2$), the dataset was standardized using Z-score normalization (zero mean, unit variance) via \texttt{StandardScaler} to ensure isotropic distance calculations in the clustering phase.

\paragraph{Supervisory Signal Generation:}
Ground truth labels for a subset of the data were derived from the catalogue of \citet{Dodd2023}, identifying 8 known substructures (e.g., GSE, Sequoia). 
To simulate a realistic semi-supervised scenario and generate robust inputs for the algorithms (Seeds for CLiMB, Must-Link/Cannot-Link for baselines), we employed a radial sampling strategy:
\begin{enumerate}
    \item For each known cluster, we calculated the centroid in the scaled feature space.
    \item Member points were sorted by their Euclidean distance to the centroid.
    \item We sampled a percentage of points (defaulting to 90\% for the main benchmark, and varying from 10\% to 100\% for sensitivity analysis) using stratified sampling along this radial distribution.
\end{enumerate}
This strategy ensures that the supervisory signals are representative of the cluster's spatial extent, providing a fair test of the algorithms' ability to generalize from core samples to peripheral members.

\subsection{Comparative Benchmark with Alternative Semi-Supervised Methods}

The selection of benchmark algorithms was guided by the intrinsic characteristics of the domain of astrophysical problems. Stellar structures in the Galactic halo, such as streams and associations, especially, traced by the relatively sparse RRLs, are often filamentary, and exhibit widely varying densities, embedded within a large, unstructured background of field stars. These properties render traditional partitional clustering algorithms, like K-Means, fundamentally unsuitable. Such methods typically require the number of clusters, \(k\), to be specified a priori, contrary to our goal of discovering novel structures, and tend to discover only convex or globular-shaped clusters, failing to capture the irregular morphologies of galactic substructures.

For these reasons, our benchmark focuses exclusively on the family of density-based clustering algorithms. These methods are uniquely suited to this problem as they can identify clusters of arbitrary shape, don't require a predefined number of clusters, and can naturally distinguish meaningful structures from low-density noise. To provide the most relevant and challenging comparison for CLiMB, we selected two alternative semi-supervised algorithms from this family, specifically chosen because they represent distinct and fundamentally different philosophies for integrating prior knowledge:

\begin{itemize}
    \item \textbf{Heuristic SS-DBSCAN:} This approach, is based on the publicly available code from Zaki\footnote{\url{https://github.com/TibaZaki/SS_DBSCAN}}, which corresponds to the method described in \citet{abdulhameed2024ss}, that modifies the cluster expansion mechanism of DBSCAN. Although the standard DBSCAN expands through any core point, this variant allows a core point to propagate the cluster label only if it satisfies a domain-specific "importance" function $\mathcal{I}(\mathbf{x})$. Based on the physical properties of accreted halo stars, we implemented $\mathcal{I}(\mathbf{x})$ to select stars that reside in the kinematic tails of the distribution. Specifically, a point is considered important if its angular momentum $L_z$ is in the bottom 15\% of the range (retrograde/low-rotation) or its total energy $E$ is in the top 15\% (high-energy). This represents a \textit{top-down, static supervision} paradigm designed to filter out background field stars.
    \item \textbf{C-DBSCAN (Constraint-based DBSCAN):} This algorithm exemplifies the classic \textit{bottom-up, constraint-based} supervision paradigm. It operates in multiple stages by first creating cautious, local micro-clusters that respect Cannot-Link constraints, and subsequently merging them based on Must-Link constraints and proximity. Our implementation was developed to follow the original algorithm described by \citet{ruiz2007c}.
\end{itemize}

By benchmarking against these two methods, we test CLiMB's performance against both a top-down, domain-guided approach and a bottom-up, constraint-based approach. The prior knowledge of the 8 known substructures was used to generate seed points for CLiMB, define the heuristic rule for SS-DBSCAN (based on extreme kinematic properties), and generate Must-Link/Cannot-Link constraints for C-DBSCAN. To ensure a fair comparison, a hyperparameter optimization was performed for each algorithm to find its optimal configuration before the final benchmark execution. (see Appendix~\ref{app:optimization} for details).

\subsection{Evaluation Methodology}
To provide a comprehensive and fair assessment, we designed an evaluation strategy that addresses the specific challenges of our astrophysical dataset. Given that our ground truth is incomplete, containing a large set of 4,465 unclassified "field stars", a single global metric is insufficient and potentially misleading. Our evaluation is therefore structured into three distinct analyses.

First, we conduct a detailed analysis of knowledge recovery. The primary goal is to measure how effectively each semi-supervised algorithm reconstructs the known substructures when provided with prior knowledge. For this, we evaluate performance exclusively on the subset of data corresponding to the 8 known clusters. We use the Adjusted Rand Index (ARI) defined by \cite{hubert1985comparing}, as the main performance score, supplemented by homogeneity and completeness as diagnostic metrics (\cite{rosenberg2007v}). These allow us to distinguish between algorithms that erroneously merge distinct substructures (low homogeneity) and those that unnecessarily fragment them (low completeness). For CLiMB, this evaluation is performed on the output of its first (constrained) phase to specifically isolate the performance of its knowledge-anchoring mechanism.

Second, we assess the quality of exploratory discovery. This analysis focuses on the new clusters identified by each algorithm from the unclassified field stars. Since no ground truth exists for these potential discoveries, we employ a qualitative and domain-based validation. We utilize phase-space diagnostic plots to verify that the newly identified clusters form coherent, dense morphologies ($E, L_z, L_\perp$) distinct from the background noise. Furthermore, we cross-reference these discoveries with recent astrophysical literature \citep{Malhan2024} to confirm their physical plausibility.

Finally, we perform a sensitivity analysis to understand how robust each algorithm is to the amount of supervision. We systematically vary the percentage of known points used as seeds (from 10\% to 100\%) and measure the resulting ARI on the known-data subset. This experiment reveals how efficiently each algorithm leverages prior knowledge, distinguishing between methods that learn monotonically and those that stagnate.

\section{Results and Discussion}
\label{sec:results}

Our experimental evaluation demonstrates that CLiMB consistently and significantly outperforms the alternative semi-supervised methods across all aspects of the analysis. The results not only establish the superiority of our proposed framework but also provide key insights into the strengths of its two-phase, adaptive-metric architecture.

\subsection{Knowledge Recovery and Baseline Comparison}

\begin{figure*}[ht!]
\centering
\includegraphics[width=0.9\textwidth]{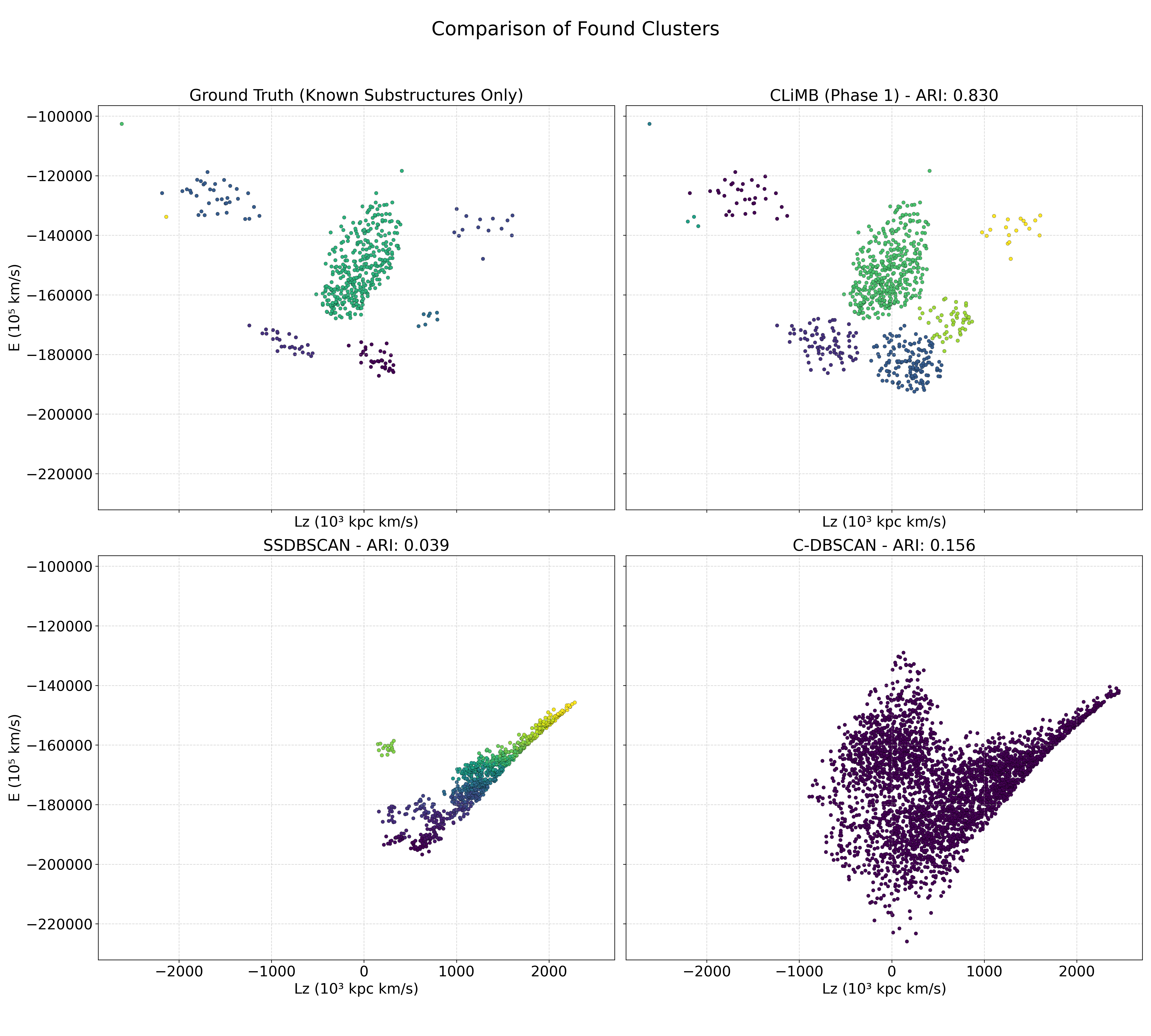}
\caption{Visual benchmark of clustering results on the known subset of structures. In all panels, distinct colors represent separate cluster assignments. Top-Left: Partial ground truth for the 8 significant substructures. Top-Right: CLiMB output (ARI: 0.829) showing accurate recovery of complex, non-convex shapes. Bottom-Left: Heuristic SS-DBSCAN (ARI: 0.040) fails to form coherent structures, fragmenting streams into noise. Bottom-Right: C-DBSCAN (ARI: 0.152) suffers from massive over-merging, collapsing distinct kinematic streams into a single macro-cluster.}
\label{fig:comparison_panel}
\end{figure*}

We first assess the ability to reconstruct the 8 known substructures using the full set of available constraints. Figure~\ref{fig:comparison_panel} presents a visual comparison between the ground truth and the final clustering outputs.

As shown in Table~\ref{tab:known_part_results}, CLiMB achieves an Adjusted Rand Index (ARI) of 0.829 (at 90\% supervision), producing a partition that closely mirrors the topological complexity of the ground truth. The high \textit{Homogeneity} score (0.953) confirms that the Mahalanobis-based anchoring effectively prevents the "leaking" of centroids into neighboring distinct structures.

In stark contrast, the baselines exhibit critical failure modes. C-DBSCAN (Figure~\ref{fig:comparison_panel}, Bottom-Right) achieves an ARI of only 0.152. The visualization reveals that its bottom-up constraint propagation is prone to ``chain-reaction'' merging in high-density regions, obliterating the boundaries between distinct stellar streams. Similarly, Heuristic SS-DBSCAN (Figure~\ref{fig:comparison_panel}, Bottom-Left) yields an ARI of 0.040, demonstrating that a single global density-heuristic is insufficient to capture the varying morphologies of galactic substructures.

This performance gap highlights the critical role of CLiMB's radial constraint. Unlike C-DBSCAN, which allows cluster centers to migrate freely based on local density bridges, often leading to 'semantic drift' where distinct structures merge via background noise, CLiMB utilises the seed centroids as physical anchors. By constraining the centroid displacement ($\tau_r$), the algorithm is forced to adapt the cluster shape (covariance) to the local data distribution without losing the kinematic identity ($E, L_z$) defined by the prior literature. This ensures that the known structures are tightly recovered, leaving the residual space clean for the subsequent exploratory phase.

\begin{table*}[ht!]
\centering
\caption{Performance on Knowledge Recovery (8 known substructures). CLiMB is compared against the final output of baselines. Bold denotes best performance.}
\label{tab:known_part_results}
\begin{tabular}{lccc}
\hline
Algorithm & ARI & Homogeneity & Completeness \\
\hline
\textbf{CLiMB} & \textbf{0.829} & \textbf{0.953} & \textbf{0.783} \\
Heuristic SS-DBSCAN    & 0.040          & 0.024          & 0.465          \\
C-DBSCAN              & 0.152          & 0.169          & 0.250          \\
\hline
\end{tabular}
\end{table*}

\subsection{Exploratory Discovery Performance}

\begin{figure*}[ht!]
\centering
\includegraphics[width=0.8\textwidth]{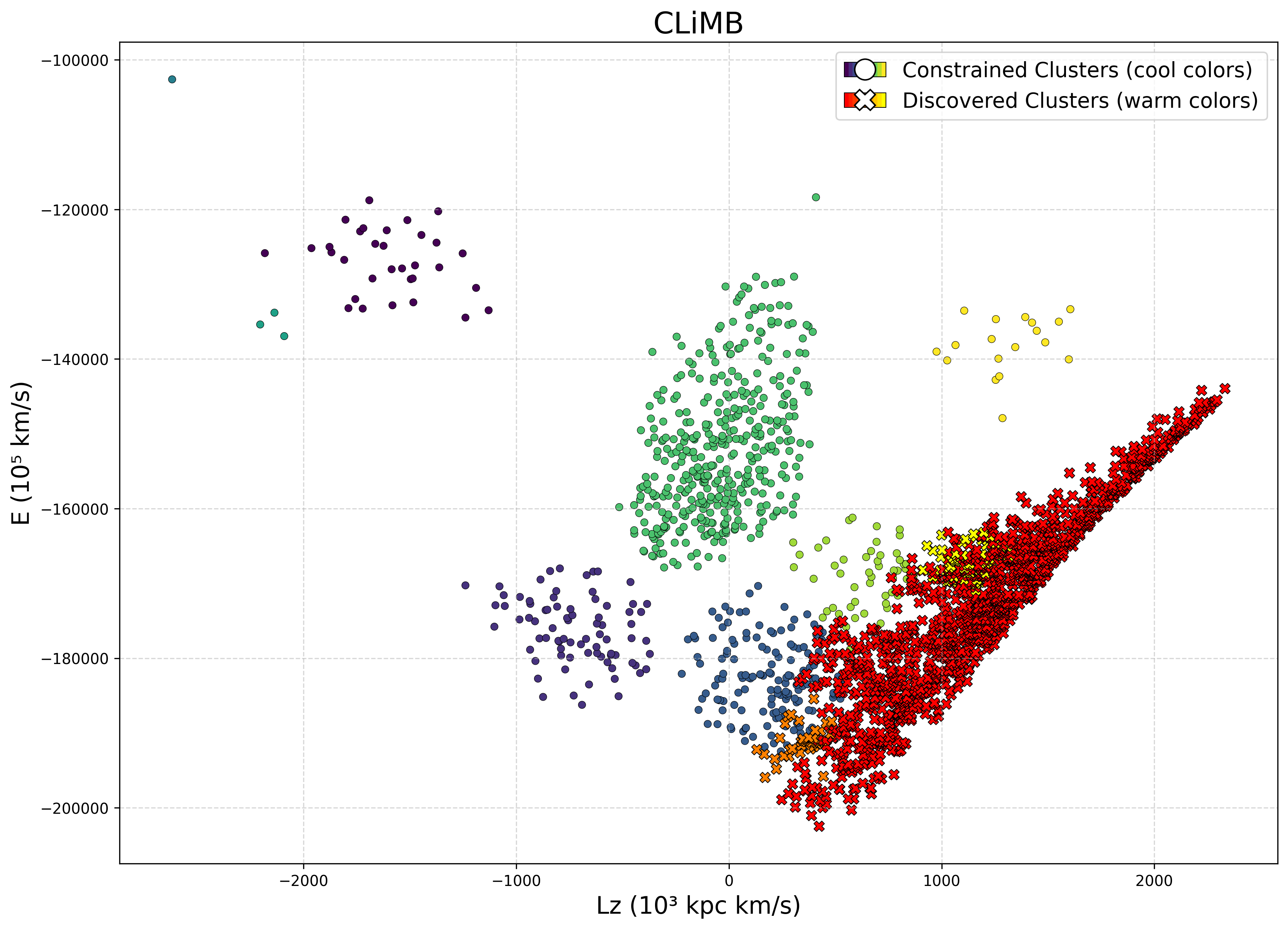}
\caption{Diagnostic plot of CLiMB's final result in the $E$-$L_z$ plane. Points are categorized by both shape and color to distinguish the algorithmic phases: circles in cool colors represent structures recovered during the constrained phase (Phase 1), while crosses in warm colors denote clusters identified during the exploratory phase (Phase 2). This diagnostic validates the detection of the \textit{Shiva} (orange) and \textit{Shakti} (yellow) structures, as well as the Galactic Disk (red), all recovered as exploratory discoveries.}
\label{fig:climb_diagnostic}
\end{figure*}

The second evaluation assessed the quality of new clusters discovered within the 4,465 out of 4,933 unclassified field stars. This test measures the algorithm's ability to perform meaningful novelty detection beyond the initial constraints. CLiMB demonstrates a robust capability in isolating latent structures. Its exploratory phase successfully identified three distinct, geometrically coherent groups in the residual data. As visually confirmed by the diagnostic plot in Figure~\ref{fig:climb_diagnostic}, these new clusters (represented by warm-colored crosses) are not mere stochastic noise; they occupy well-defined, high-density regions of the ($E$-$L_z$) phase space and are clearly separable from the constrained clusters (cool-colored circles). Subsequent domain analysis confirms that these groups correspond to the Galactic Disk and the recently hypothesized \textit{Shiva} and \textit{Shakti} substructures \citep{Malhan2024}, validating the physical significance of the resulting partition. In contrast, the baselines failed to produce comparable insights. As seen in the comparison panel (Figure~\ref{fig:comparison_panel}), C-DBSCAN failed to identify these novelties, merging the majority of the field stars into the known structures or classifying them as noise. Heuristic SS-DBSCAN produced fragmented, incoherent groupings that do not align with known kinematic morphologies. This confirms that without the "cleaning" effect of CLiMB's first phase, standard density-based methods struggle to distinguish subtle novel signals from the dominant background noise.

\subsection{Overall Performance and Sensitivity to Supervision}

The final analysis considers the overall performance and robustness. The "Global Meaningful ARI" confirms CLiMB's lead with a score of 0.829, compared to 0.040 for SS-DBSCAN and 0.152 for C-DBSCAN. 

\begin{figure*}[ht!]
\centering
\includegraphics[width=0.8\textwidth]{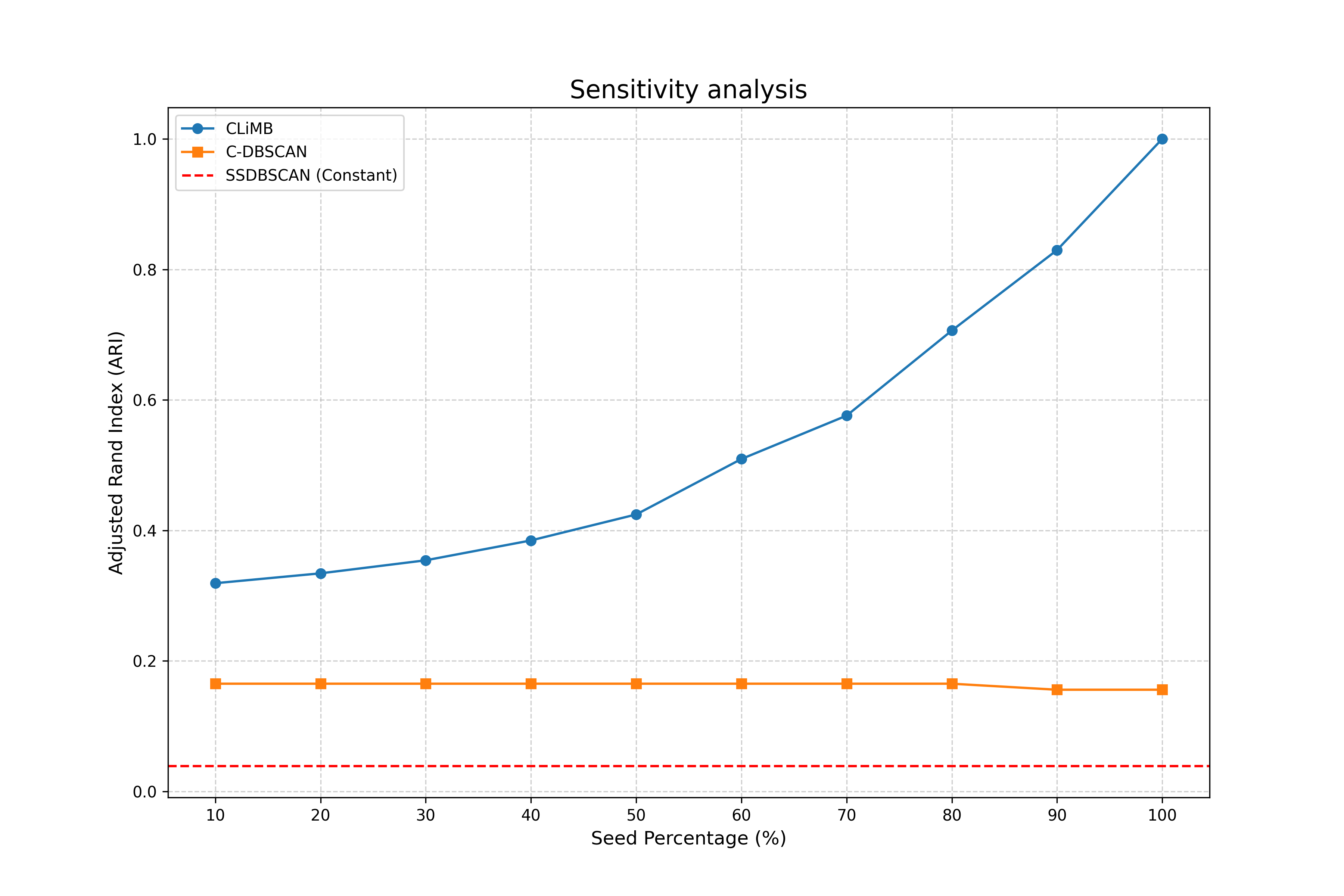} 
\caption{Sensitivity analysis showing the Adjusted Rand Index (ARI) on knowledge recovery as a function of the percentage of prior knowledge (10\% to 100\%). CLiMB (blue) demonstrates monotonic learning and high performance even with limited supervision, whereas baselines (C-DBSCAN and SSDBSCAN in orange and red) stagnate.}
\label{fig:sensitivity_plot}
\end{figure*}

Furthermore, the sensitivity analysis, shown in Figure~\ref{fig:sensitivity_plot}, highlights CLiMB's remarkable robustness and efficiency in learning from prior knowledge. A crucial divergence in algorithmic behavior is observed: while C-DBSCAN and SS-DBSCAN exhibit stagnant performance regardless of the amount of supervision (flat lines), CLiMB demonstrates a strict monotonic improvement. Notably, even with sparse prior knowledge (10\% of seeds), CLiMB achieves an ARI ($\approx 0.32$) that doubles the performance of the best baseline, scaling up to 0.829 as more information is provided.

This demonstrates that CLiMB's architecture effectively translates extensive domain knowledge into topological accuracy without becoming over-constrained, a key advantage over traditional methods that fail to leverage additional labels.

\subsection{Limitations and Operational Trade-offs}

While CLiMB demonstrates superior accuracy and discovery capabilities, these advantages come with specific operational trade-offs regarding configuration and complexity. 

First, the dual-phase architecture inherently expands the hyperparameter space compared to monolithic algorithms. Beyond the standard density parameters ($\epsilon, \text{MinPts}$) required for the exploratory phase, the constrained phase necessitates the definition of specific anchoring thresholds, namely density ($\tau_{\rho}$), distance ($\tau_{d}$), and radial drift ($\tau_{r}$). While this offers granular control, it requires a hierarchical optimization strategy as detailed in Appendix~\ref{app:optimization}. Second, as analyzed in Section~\ref{subsec:complexity}, the reliance on Gaussian Kernel Density Estimation for the initial filtering introduces a pre-computation cost of $O(n^2 d)$ in the naive implementation. While the iterative assignment phase is efficient, this initialization step represents the computational bottleneck for large datasets, necessitating the use of approximation structures (e.g., kd-trees) for samples exceeding millions of points. Finally, the architectural decoupling introduces a sequential dependency: the quality of the Novelty Detection is strictly conditional on the successful anchoring of the first phase. If the constrained constraints are too lax, novelties may be absorbed into known clusters; if too strict, known points may leak into the residual space, contaminating the discovery process.

\section{Conclusions}
\label{sec:conclusions}

In this work, we introduced CLiMB, a domain-informed novelty detection framework designed to resolve the fundamental challenge in semi-supervised clustering: balancing the adherence to prior knowledge with the flexibility to discover unanticipated structures. By architecturally decoupling the problem into a constrained anchoring phase (\textit{K-Bound}) and a density-based exploratory phase, CLiMB overcomes the ``representativeness'' assumption that limits traditional constraint-based methods. This separation allows  practitioners to rigidly enforce known physical or biological constraints where data is well-characterized, while retaining the mathematical freedom to detect non-convex, topological anomalies in the unlabelled residual data.

The practical efficacy of this approach was demonstrated through a challenging application in Galactic archaeology using \textit{Gaia} DR3 data. CLiMB achieved a significant performance advantage over existing semi-supervised baselines, demonstrating robust learnability by scaling to an ARI of 0.829, whereas competitor methods stagnated below 0.20 regardless of the supervision level. Beyond metric performance, the framework facilitated genuine scientific discovery: the algorithm's ability to "clean" the feature space in the first phase allowed for the isolation of the \textit{Shiva} and \textit{Shakti} substructures, along with the Galactic Disk, from the residual background. These results underscore CLiMB's potential as a hypothesis generation tool for high-dimensional scientific domains where valid signals are sparse and the catalogue of known phenomena is incomplete.

Despite these advancements, the current framework exhibits specific limitations that outline the path for future research. First, the dual-phase architecture introduces a higher dimensionality in the hyperparameter space compared to monolithic algorithms; while partially mitigated by our hierarchical optimization strategy, the dependence on user-defined thresholds ($\tau_{\rho}, \tau_{d}, \tau_{r}$) assumes some domain intuition regarding cluster scales. Second, the computational complexity of the initial Gaussian density estimation ($O(n^2 d)$) poses scalability challenges for datasets exceeding millions of points, necessitating the future integration of approximate nearest-neighbor methods (e.g., kd-trees). Finally, the sequential nature of the algorithm implies that the exploratory phase is sensitive to the quality of the initial constrained assignment; future iterations could explore a joint optimization objective that refines both phases simultaneously to minimize leakage.

\section*{Acknowledgements}{Support to this study has been provided by INAF Mini-Grant (PI: Tatiana Muraveva), by the Agenzia Spaziale Italiana (ASI) through contract ASI 2018-24-HH.0, its Addendum 2018-24-HH.1-2022 and contract ASI 2025-10-H.00, and by Premiale 2015, MIning The Cosmos - Big Data and Innovative Italian Technology for Frontiers Astrophysics and Cosmology (MITiC; P.I.B.Garilli).
This work uses data from the European Space Agency mission {\it Gaia} (\url{https://www.cosmos.esa.int/gaia}), processed by the {\it Gaia} Data Processing and Analysis Consortium (DPAC; \url{https://www.cosmos.esa.int/web/gaia/dpac/consortium}). Funding for the DPAC has been provided by national institutions, in particular the institutions participating in the {\it Gaia} Multilateral Agreement.}

\appendix

\section{Hyperparameter Optimization}\label{app:optimization}

To ensure a fair and robust comparison, a hyperparameter optimization process was conducted for each algorithm prior to the final benchmark execution. The primary objective of this process was to identify the optimal parameter configuration for each method on our specific dataset. For the density-based competitors (Heuristic SS-DBSCAN and C-DBSCAN), we performed a grid search over a range of values for \texttt{eps} and \texttt{min\_samples}, selecting the combination that maximized the "Global ARI" as described in the main text.

For our proposed algorithm, CLiMB, we employed the Mahalanobis distance metric for the constrained phase to naturally model the ellipsoidal correlations typical of kinematic streams ($E$-$L_z$). We then used a hierarchical, two-stage optimization strategy:

\begin{enumerate}
    \item \textbf{Phase 1 (Constrained) Optimization:} We first optimized the parameters governing the knowledge-recovery phase (\texttt{density\_threshold}, \texttt{distance\_threshold}, etc.) by maximizing the ARI calculated exclusively on the subset of 8 known substructures.
    \item \textbf{Phase 2 (Exploratory) Optimization:} Using the optimal parameters from the first stage, we then optimized the exploratory parameters (\texttt{eps}, \texttt{min\_samples}) by minimizing the Davies-Bouldin Index (DBI) \citep{davies2009cluster} on the clusters discovered within the unclassified data.
\end{enumerate}

This hierarchical approach ensures that each phase of CLiMB is fine-tuned for its specific task. The final, optimal hyperparameters used for the benchmark analysis presented in this paper are detailed in Table~\ref{tab:hyperparameters}. The complete code for the optimization routines, along with the scripts to reproduce all results, is publicly available in our GitHub repository\footnote{\url{github.com/LorenzoMonti/RRLs-Clustering-Benchmark}}.

\begin{table*}[t!]
\centering
\caption{Optimal Hyperparameters Used in the Benchmark.}
\label{tab:hyperparameters}
\begin{tabular}{llc}
\hline
Algorithm & Parameter & Optimal Value \\
\hline
\rule{0pt}{3ex}
\textbf{CLiMB} & \multicolumn{2}{l}{\textit{Phase 1 (Constrained)}} \\
& \texttt{Metric} & Mahalanobis \\
& \texttt{density\_threshold} & 0.005 \\
& \texttt{distance\_threshold} & 0.5 \\
& \texttt{radial\_threshold} & 0.1 \\
& \texttt{convergence\_tolerance} & 0.01 \\
\rule{0pt}{3ex}
& \multicolumn{2}{l}{\textit{Phase 2 (Exploratory)}} \\
& \texttt{eps} & 0.190 \\
& \texttt{min\_samples} & 24 \\
\hline
\rule{0pt}{3ex}
\textbf{Heuristic SS-DBSCAN} & \texttt{eps} & 0.15 \\
& \texttt{min\_samples} & 16 \\
\hline
\rule{0pt}{3ex}
\textbf{C-DBSCAN} & \texttt{eps} & 0.24 \\
& \texttt{min\_pts} & 16 \\
\hline
\end{tabular}
\end{table*}

\bibliographystyle{elsarticle-harv} 
\bibliography{bib}


\end{document}